\documentstyle[epsfig]{aipproc}

\newcommand{\NP}[1]{Nucl. Phys.\ {\bf #1}\ }
\newcommand{\PL}[1]{Phys. Lett.\ {\bf #1}\ }

\newcommand{\PR}[1]{Phys. Rev.\ {\bf #1}\ }
\newcommand{\PRL}[1]{Phys. Rev. Lett.\ {\bf #1}\ }

\newcommand{\DE}{\Delta}

\newcommand{\tco}{\tilde\chi_1 }
\newcommand{\tct}{\tilde\chi_2 }

\newcommand{\sulu}{$SU(2)_L\times U(1)_Y$}

\newcommand{\matr}{\left( \begin{array}}
\newcommand{\ematr}{\end{array} \right)}

\begin{document}

\title{Phenomenology of SUSY-models with spontaneously broken
R-parity\thanks{Talk given in Beyond the Standard Model V in Balholm,
Norway. Presented by K. Puolam\"{a}ki.}}

\author{K. Huitu$^*$, J. Maalampi$^{\dagger}$ and
K. Puolam\"{a}ki$^*$}

\address{$^*$Helsinki Institute of Physics, P.O.Box 9, FIN-00014
University of Helsinki\\$^{\dagger}$Department of Physics, Theoretical
Physics Division, P.O.Box 9,\\FIN-00014 University of Helsinki}

\maketitle

\begin{abstract}

We review the consequences of spontaneously broken R-parity in present
and planned lepton-lepton colliders. In the left-right models the
R-parity, $R=(-1)^{3(B-L)+2S}$, is preserved due to the gauge
symmetry, but it must be spontaneously broken in order to the scalar
spectrum to be physically consistent. The spontaneous breaking is
generated via a non-vanishing VEV of at least one of the sneutrinos,
which necessarily means non-conservation of lepton number $L$. The
R-parity violating couplings are parametrized in terms of mixing
angles, whose values depend on model parameters. Combined with the
constraints derived from low-energy measurements this yields allowed
ranges for various R-parity breaking couplings. The R-parity breaking
allows for the processes in which a single chargino or neutralino is
produced, subsequently decaying at the interaction point to
non-supersymmetric particles.

\end{abstract}

\section*{Introduction} 

In minimal supersymmetric model \cite{SUSY} the gauge symmetry and
supersymmetry allow three point interactions and bilinear terms that
violate either baryon- or lepton number. These interaction terms are
restricted by a plethora of as yet unobserved low energy phenomena
(see e.g. \cite{chaichian96}). If both baryon and lepton number
violation are realized, the proton will decay fast \cite{proton}. To
cure these problems one usually assumes so-called R-parity, defined as
$(-1)^{3(B-L)+2S}$, where $B$ is a baryon number, $L$ a lepton number
and $S$ spin of the matter field, to be a conserved global symmetry
\cite{farrar78,dimopoulos81,martin96}.

The assumption of R-parity symmetry is not necessary for the internal
consistency of the model. Furthermore, the global symmetries at
electroweak scale could be violated by Planck scale effects
\cite{giddings88}. A more natural way would be to have the matter
parity as a gauge symmetry. The simplest physically consistent model
to accomplish this is the supersymmetric left-right model (SLRM) based
on a gauge group $SU(2)_L \times SU(2)_R \times U(1)_{B-L}$
\cite{francis91}. In SLRM the lepton superfields, and similarily as
quark fields, are arranged in $SU(2)_{L,R}$ doublets $L$ and $L^c$.

The matter parity is a discrete subgroup of $U(1)_{B-L}$ gauge
group. It can be shown that in $U(1)_{em}$ preserving minimum at least
one sneutrino has a non-vanishing VEV $\langle \nu^c_j \rangle =
\sigma_R \ne 0$, which is at least of the order of the breaking scale
of the left-right symmetry or supersymmetry
\cite{kuchimanchi93,huitu95}. Lepton number --- and consequently
R-parity --- is thus spontaneusly violated. We have considered the
phenomenological consequences of this model in \cite{huitu97}. Here we
will review the main results of our study.

\section*{Supersymmetric left-right model}

The minimal Higgs content of the left-right susy model is two
$SU(2)_R$ triplets and two $SU(2)_L\times SU(2)_R$ bidoublets:
\begin{eqnarray}
\label{higgses}
 \DE =\matr{c} \DE^{0} \\ \DE^{-}\\ \DE^{--} \ematr \sim ({\bf
  1,3,}-2), \,\,\,\,\,\,\,\,\,\,\,\, && \delta =\matr{c} \delta^{++}
  \\ \delta^{+} \\ \delta^{0} \ematr \sim ({\bf 1,3,}2), \nonumber \\
  \phi =\matr{cc} \phi_1^0& \phi_1^+\\ \phi_2^-& \phi_2^0 \ematr \sim
  ({\bf 2,2,}0), \,\,\,\,\,\,\,\,\,\,\,\, && \varphi =\matr{cc}
  \varphi_1^0& \varphi_1^+\\ \varphi_2^-& \varphi_2^0 \ematr \sim
  ({\bf 2,2,}0). \nonumber \end{eqnarray} Two triplet superfields are
  required in order to cancel $U(1)_{B-L}$ anomalies due to higgsino
  loops and two Higgs bidoublets are required to obtain a non-trivial
  CKM quark mixing matrix.

The spontaneous symmetry breaking takes place at two separate
scales. At the first stage $SU(2)_L \times SU(2)_R \times U(1)_{B-L}$
is broken to \sulu by triplet Higgs and sneutrino VEV's:
\begin{eqnarray} \langle \Delta^0 \rangle = v_\Delta \mbox{ , }
\langle \delta^0 \rangle = v_\delta \mbox{ , } \langle \nu^c
\rangle = \sigma_R \mbox{ .}  \nonumber
\end{eqnarray}
This breaking sets the massscale of the right-handed gauge bosons (one
could have $m_{W_R}= {\cal{O}} (1 ... 10)$ TeV). The mass of the
charged boson $W_R$ is given by \cite{huitu95}
\begin{eqnarray}
\nonumber
m_{W_R}^2 = g_R^2 \left(v_\Delta^2+v_\delta^2+\frac{1}{2} \sigma_R^2
\right) \mbox{ .}
\nonumber
\end{eqnarray}
Experimentally $m_{W_R}>650$ GeV \cite{tevatron}. The standard model
symmetry group \sulu is broken to $U(1)_{em}$ by bidoublet VEV's 
\begin{eqnarray}
\langle \Phi^0_1 \rangle = \kappa_1 \mbox{ , } \langle \varphi^0_2
\rangle = \kappa_2 \mbox{ ; } \tan \beta =
\frac{\kappa_2}{\kappa_1}\mbox{ .}  \nonumber
\end{eqnarray}
The connection of this breaking scale to the mass of the ordinary
left-handed gauge boson $W_L$ is given by
\begin{eqnarray}
m_{W_L}^2 = \frac{1}{2} g_L^2 \left( \kappa_1^2+\kappa_2^2 \right) =
\left(80\mbox{GeV}\right)^2 \mbox{ .}
\nonumber
\end{eqnarray}

As a result of the $R$-parity and lepton number breaking, arising when
$\langle \nu ^c \rangle = \sigma_R \ne 0$, leptons ($\nu$,$l$) mix
with gauginos ($\lambda$) abd higgsinos ($\tilde h$). The physical
neutralino and chargino states most generally are thus the following
kind of superpositions:
\begin{eqnarray} \tilde \chi^0_i & = & a_{ij}^{(0)} \lambda^0_j +
b_{ij}^{(0)}\tilde h^0_j +c_{ij}^{(0)} \nu_j {\mbox{   ,}} \nonumber
\\ \tilde \chi^\pm_i &=& a_{ij}^{(\pm)} \lambda^\pm_j + b_{ij}^{(\pm)}
\tilde h^\pm_j + c_{ij}^{(\pm)} l^\pm_j {\mbox{  .}}  \nonumber
\end{eqnarray}

In contrast with lepton number, the baryon number is conserved, since
spontaneous baryon number violation would indicate a presence of a
non-vanishing VEV of a squark field, whic would break the
$SU(3)_C\times U(1)_{em}$ gauge symmetry.

In principle every sneutrino could have a VEV. However, either of the
VEV's for electron and muon sneutrinos $\langle \nu_{e}^c \rangle $
and $\langle \nu_{\mu}^c \rangle$ should be small or vanish, since
otherwise there would be a tree-level amplitude for lepton family
number violating decay $\mu \rightarrow 3e$, leading to a branching
ratio that contradicts with the experimental bound $\mbox{B}\left( \mu
\rightarrow 3e \right) < 1.0 \times 10^{-12}$ \cite{PDG}.

In SLRM the physical leptons are chargino mass eigenstates composed of
gauginos, Higgsinos and leptons. Therefore their couplings to
$Z$-boson will deviate from their normal form. However, the deviations
are not large in general as the components which form the dominant
part of the mass eigenstates of charginos and neutralinos
corresponding to the SM leptons have all approximately the SM
couplings with $SU(2)_L$ gauge bosons.  This is so due to the symmetry
breaking structure: the states that diagonalize the Lagrangian after
the $SU(2)_R \times U(1)_{B-L}$ breaking have definite $SU(2)_L \times
U(1)_Y$ quantum numbers and couplings. The subsequent breakdown to
$U(1)_{em}$ cause only small deviations to the composition of the
physical states, characterized by $(m_{W_L} \mbox{ or }
m_\tau)/m_{W_R}$ or $m_\tau/m_{susy}$, where $m_{susy}$ is a general
susy scale.

We assume that for simplicity only one sneutrino, namely
${\nu}_\tau^c$, will have a non-vanishing VEV.

\section*{Testing the model in electron-positron colliders}

We have considered in \cite{huitu97} two representative models. In the
first model the soft gaugino mass terms are about 100 GeV. The
lightest supersymmetric particles are almost mass-degenerate
neutralino ($m_{\tct^0} \simeq 93$ GeV) and chargino ($m_{\tct^\pm}
\simeq 94$ GeV) both almost purely gaugino $\lambda_L$ states.

\begin{figure}
\centerline{\epsfig{file=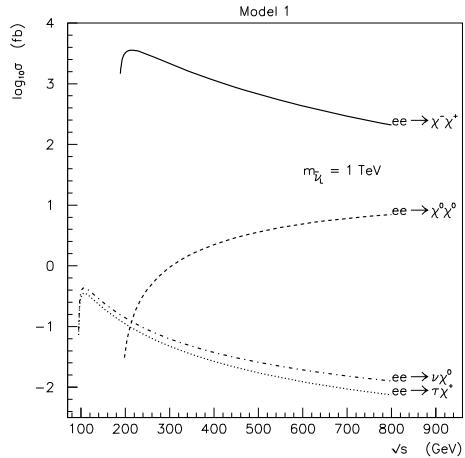,height=7.cm,width=7.cm}}
\caption{The production cross section of single or
pair produced neutralinos and charginos as a function of center of
mass energy in the Model 1 with $m_{\tilde e_L}=m_{\nu_{eL}}=1$
TeV.}
\label{F:plot1b} 
\end{figure}

In Fig. \ref{F:plot1b} we have plotted the cross sections for single
production of the lightest charginos and neutralinos via reactions
$e^+e^-\to \tct^+\tau^-$ and $e^+e^-\to \tct^0\nu_{\tau}$ and for the
pair production via the reactions $e^+e^-\to \tct^+\tct^-$ and
$e^+e^-\to \tct^0\tct^0$ as a function of the center of mass energy
$\sqrt{s}$. These processes occur via a gauge boson exchange in
s-channel and via a t-channel sneutrino or selectron exchange.

One can test the model by studying the production and decays of
charginos and neutralinos in electron-positron colliders. LEP 2 is
currently running at 160--192 GeV center of mass energies and it will
collect about 0.5 $fb^{-1}$ total integrated luminosity by the end of
the milleneum. The linear collider is planned to operate initially at
center of mass energy $\sqrt{s}=380$ GeV with an annual luminosity of
${\cal{L}}_{year}=10$ $fb^{-1}$ and to be gradually upgraded to 1.6
TeV with ${\cal{L}}_{year}=200 fb^{-1}$ annual luminosity
\cite{wiik96}.

From these characteristic numbers we can conclude that the single
production processes are in Model 1 too suppressed to be visible in
LEP 2 or at planned linear colliders. The pair production, when
kinematically allowed, would in turn contribute hundreds of reactions.

In Model 1 the neutralino decays at the interaction point to the tau
lepton and the ordinary $W$-boson with the decay width of
$\Gamma(\tct^0\rightarrow \tau+W) \simeq 10$ keV. The chargino can
decay either via gauge or Higgs boson, decay widths being $\Gamma
(\tct^+\rightarrow \nu_\tau W) \simeq 20$ keV and $\Gamma
(\tct^+\rightarrow \tau^+H_1^0) \simeq 10$ keV. In the latter case the
Higgs boson will further decay, mainly to a pair of bottom quarks.

\begin{figure}
\centerline{\epsfig{file=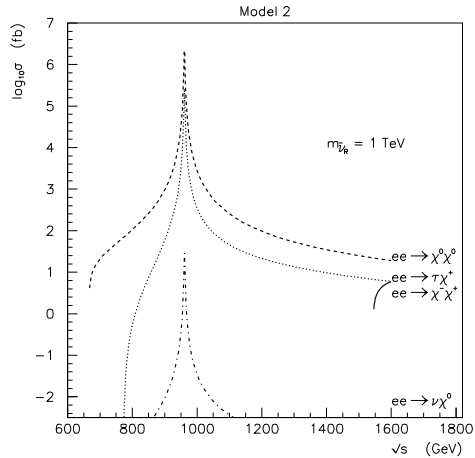,height=7.cm,width=7.cm}}
\caption{\label{M2cs} The production cross section of single or pair
produced neutralinos and charginos as a function of center of mass
energy in Model 2 with $m_{\tilde e_R}=m_{\nu_{eR}}=1$ TeV. }
\label{F:plot2b} 
\end{figure}

In Model 2 all soft masses are of the order 1 TeV. An interesting
feature of this model is the existence of a relatively light doubly
charged Higgs boson $H^{++}$ ($m_{H^{++}} =334 GeV$). The lightest
supersymmetric chargino ($m_{\tct^\pm}=772$ GeV) and neutralino
($m_{\tct^0}=333$ GeV) pair production would be visible in linear
collider, whenever kinematically allowed. In this particular model a
relatively light extra $Z$-boson, $Z_2$, ($m_{Z_2} \simeq 1$ TeV) also
contributes and allows for measureable production cross section for
single chargino $\tct^+$ and $\tco^- \sim \tau^-$, as shown in
Fig. \ref{F:plot2b}.

The neutralino would decay mainly to a neutrino and the lightest Higgs
boson $\Gamma (\tct^0\rightarrow \nu_\tau H_1^0) \simeq 100$keV. The
chargino would decay predominantly to the tau lepton and the doubly
charged Higgs boson, which would further decay to same sign lepton
pair. The decay widths for the chargino are $\Gamma( \tct^+
\rightarrow \tau^- H^{++} ) \simeq 90$ MeV and $\Gamma
(\tct^+\rightarrow \tau^+Z) \simeq 3$ MeV.

\section*{Summary}

In contrast with the supersymmetric standard model, in extended
models, e. g. in the left-right model, R-parity symmetry can be a
subgroup of the gauge symmetry, and is spontaneously broken. The
spontaneous symmetry breaking mechanism then uniquely determines the
R-parity breaking couplings.

As a result of R-parity breaking charginos and neutralinos can in
principle be singly produced. We have studied the prospects to measure
this production at LEP 2 and linear collider and we have compared it
with the pair production.

\end{document}